\documentclass[authoryear,preprint,review,12pt]{elsarticle}

\usepackage{graphicx}
\usepackage{subfigure}
\usepackage{natbib} 
\bibpunct{[}{]}{;}{n}{,}{,} 
\usepackage{url}

\usepackage{color}
\usepackage{xspace}  
\newcommand{\FLASHthree}{FLASH3\xspace}

\journal{Parallel Computing}

\begin{document}

\begin{frontmatter}

\title {Extensible Component Based Architecture for FLASH, A Massively Parallel, Multiphysics Simulation Code}

\author[1]{Anshu~Dubey\corref{cor}}
\ead{dubey@flash.uchicago.edu}

\author[2]{Katie Antypas}
\author[3]{Murali K. Ganapathy}
\author[1]{Lynn B. Reid}
\author[5]{Katherine Riley}
\author[5]{Dan Sheeler}
\author[5]{Andrew Siegel}
\author[1]{Klaus Weide}
\cortext[cor]{Corresponding author}
\address[1]{ASC/Flash Center, The University of Chicago, 5640 S. Ellis Ave, Chicago, IL 60637}
\address[2]{Lawrence Berkeley National Laboratory, 1 Cyclotron Road, Berkeley, CA 94720}
\address[3]{Google Inc.}
\address[5]{Argonne National Laboratory, 9700 S. Cass Ave, Argonne, IL, 60439}

\begin{abstract}

FLASH is a publicly available high performance application code which
has evolved into a modular, extensible software system from a
collection of unconnected legacy codes.
FLASH has been
successful because its capabilities have been driven by the needs of
scientific applications, without compromising maintainability,
performance, and usability.
In its newest incarnation, \FLASHthree
consists of inter-operable modules that can be combined to
generate different applications.
The FLASH architecture allows arbitrarily
many alternative implementations of its
components to co-exist and interchange
with each other, resulting in
greater flexibility. Further, a simple and elegant mechanism
exists for customization of code functionality  without
the need to modify the core implementation of
the source.
A built-in unit test framework providing verifiability,
combined with a rigorous software maintenance process,
allow the code to operate simultaneously in the dual mode of production and
development.
In this paper we describe the \FLASHthree
architecture, with emphasis on solutions to the more challenging
conflicts arising from solver complexity, portable performance
requirements, and legacy codes. We also include results from user
surveys conducted in 2005 and 2007, which highlight the success of the
code.

\end{abstract}


\end{frontmatter}

\section{Introduction}
The ASC/Flash Center at the University of Chicago has developed a
public domain astrophysics application code,
FLASH~\citep{Fryxell2000,Calder2002a}.  FLASH is component-based,
parallel, and portable, and has a proven ability to scale to tens of
thousands of processors.  The FLASH code was developed under contract
with the Department of Energy ASC/Alliance Program. It is available to
external users through a cost-free licensing agreement. Approved users
may download the source code and make local modifications,
but may not redistribute the code. FLASH is the
flagship Computer Science product of the Flash Center, resulting from
over 10 years of research and development.  One of the mandates of the
Flash Center was the delivery of a parallel, scalable, and
highly-capable community code for astrophysics. Motivation for the code
effort lay in the increasing complexity of astrophysical
simulations.  The traditional academic approach of developing
numerical software in piecemeal was deemed inadequate to meet the science
needs.

Another aim of the Flash Center was to shift the paradigm of
theoretical research towards working in multidisciplinary teams with
scientific codes that are developed with modern software practices
prevalent in the commercial world. The FLASH code has now reached a level
of maturity where it has a large number of users, more than 80\% external to the
University of Chicago.  Moreover, it also has a substantial number of
external code contributors.  The number of requests for download, and
the number of publications using the FLASH code, have grown
superlinearly in recent years (see Section~\ref{sec:UserSurvey}).
This success was achieved by
carefully balancing the often conflicting requirements of physics,
software engineering, portability, and performance.

From its inception, FLASH has simultaneously been in development and
in production mode. Its evolution into a modern component-based code has
taken a path very different from that of most scientific computing
frameworks such as Chombo, SAMRAI, CACTUS, and POOMA
\citep{chombo2009,wissink2000,samrai2008,hornung2002,Hornung2006,Ko2005,Reynders1996,Oldham2002}.
Those efforts developed the framework first, followed by the
addition of solvers and other capabilities. An alternative path taken
by scientific application codes such as Enzo, SWMF, and Athena
\citep{O'Shea2005,toth2005,Gardiner2005} is to grow into a large application from
smaller solvers and applications. Both models of development have
their advantages and disadvantages: codes initialized with frameworks
have superior modularity and maintainability, while codes begun with
solvers generally deliver better performance for their target
applications. FLASH straddles both approaches.

In the first released version, the development followed the solvers-first
model, but later versions place more emphasis on modularity,
maintainability, and extensibility. The outcome of this duality in
development is that FLASH has more capabilities and customizability,
and it reaches a much wider community than most scientific application
codes.
FLASH has gained wide usage because the capabilities of the code have
been driven by physics, while its architecture is driven by
extensibility and maintainability.  The addition of new solvers to
FLASH is almost always dictated by the needs of users' applications.
The solvers for multiphysics applications tend to put severe strain
on any modern object-oriented software design.
Lateral data movement is normally required
between different solvers and functional units, which makes resolving
data ownership and maintaining encapsulation especially challenging.
Also, many of the
core physics solvers are legacy third-party software written
in Fortran, which are rarely modular. While modularity, flexibility,
and extensibility are some of the primary guiding principles in the
code architecture design, these goals often conflict with the equally
important considerations of efficiency and performance.  Additionally,
since high performance platforms usually have a relatively short
lifespan, the need for performance portability  places even more
constraints on the design process. Achieving a
balance between these conflicting goals while retaining the very
complex multiphysics capabilities has been the biggest
contributor to the widespread acceptance of the FLASH code.

The FLASH model of development and architecture is informed by the literature
from the common component architecture effort
\citep{hovland2003,armstrong2006}.
Since the project's inception, FLASH has undergone two major revisions,
both of which included significant architectural and capabilities
improvements. FLASH has always striven for a component-based
architecture, but this goal was not realized in the first version because
of a strong emphasis on producing early scientific results using
legacy codes. However, foundations for a component-based architecture
were firmly laid in the first version FLASH1.6 \citep{Fryxell2000} by
providing wrappers on all the solvers and minimizing lateral
communication between different solvers. The second generation versions, FLASH2.0
-- FLASH2.5, built upon this foundation by addressing data ownership
and access, resulting in a centralized data management
approach. Finally, the current version, \FLASHthree, has realized a true
component-based architecture with decentralized data management, clean
interfaces, and encapsulation of functional units. \FLASHthree also has
well-defined rules for inheritance within a unit and for interactions
such as data communication between units.
Further discussion of architecture changes over revisions is provided in
\citet{antypas2006}.

This latest release contains over 380,000 lines of code, with over
138,000 additional lines of comments. The core of the FLASH
code is written in Fortran90, with input/output interfaces provided in
C. Initially Fortran was chosen because the legacy computational kernels were
written in Fortran, whose interoperability with object-oriented
languages can be memory inefficient and unportable.  In addition,
experience with system software limitations on various
supercomputers demonstrated the wisdom of avoiding complex features
such as dynamic linking in the build process.
The choice of Fortran does affect the architecture: instead of
depending upon the programming language to enforce modular
implementation, FLASH must rely upon a combination of the Unix directory
structure and several scripts to maintain modularity (see
Figure~\ref{fig:units}).
However, lack of strong checking by the language can also be advantageous
because it discourages complexity in the design.
In addition, the ``primitive''
features of Fortran allow developers to sometimes accelerate debugging by
temporarily bypassing the architecture to give direct access to
data structures.

More than 35 developers and researchers have
contributed to all versions  of the FLASH code.
During the past 10 years, over 80 person-years of effort have
built the code and its scientific algorithms.
As the complexity of the code and the number of developers have grown,
code verification and management of the software development process
have become increasingly important to the success of the project. The
\FLASHthree distribution now includes a unit test framework and its
own test-suite, called FlashTest, which can be used for professional
regression testing.

In this paper we describe the \FLASHthree architecture, with emphasis on
solutions to the more challenging conflicts arising from solver
complexity, portable performance requirements, and legacy codes. We
also include results from user surveys conducted in 2005 and 2007, indicating
how the architecture choices have led to
the widespread acceptance of the FLASH code.

\section{Architecture Cornerstones}
\label{sec:Overview}

FLASH is not a monolithic application code; instead, it should be
viewed as a collection of components that are selectively grouped to
form various applications. Users specify which components should be
included in a simulation,  define a rough
discretization/parallelization layout, and  assign their own initial
conditions, boundary conditions, and problem setup to create a unique
application executable.
In FLASH terminology, a component that
implements an exclusive portion of the code's functionality is called
a {\em unit}.  A typical FLASH simulation requires a proper subset
of the units available in the code.  Thus, it is important to
distinguish between the entire FLASH source code and a given FLASH
application.

The FLASH architecture is defined by four cornerstones: unit,
configuration layer, data management, and interaction
between units. Here we describe the four cornerstones briefly.

\subsection{Unit}
\label{sec:Unit}
A FLASH unit provides well-defined functionality and conforms to a
structure that facilitates its interactions with other units. A unit
can have interchangeable implementations of varying complexity, as
well as subunits that provide subsets of the unit's functionality. Each
unit defines its Application Programming Interface (API), a collection
of routines through which other units can interact with it. Units must
provide a null implementation for every routine in their API. This
feature permits an application to easily exclude a unit without the
need to modify code elsewhere. For example, the input/output
unit can be easily turned on and off for testing purposes,
by linking with the null implementations.

FLASH units can be broadly classified into five functionally distinct
categories: infrastructure, physics, driver, monitoring, and simulation.
This categorization is meant to clarify the role of different classes
of units in a simulation, rather than any architectural differences
among them. In terms of organization, and their treatment by the
configuration tool, all units follow the same rules,
except the IO and the Simulation units, described in
Sections~\ref{sec:InfrastructureUnit} and \ref{sec:SimulationUnit}.
The infra\-structure category includes the units responsible for
housekeeping tasks such as the management of runtime parameters, the
handling of input and output to and from the code, and the
administration of the solution mesh.
Units of this type are discussed further in Section~\ref{sec:InfrastructureUnit}.
Units in the physics category implement algorithms to solve the equations describing
specific physical phenomena, and include units such as hydrodynamics,
equations of state, and gravity.  These units constitute the core of the FLASH solution 
capabilities.
The Driver unit implements
the time advancement methods, initializes and finalizes the
application, and controls most of the interaction between units
included in a simulation.
Because control of the simulation is implemented by the Driver
unit, it interacts the most with other individual units (see Section~\ref{sec:Interactions} for more detail).
The monitoring units track the
progress and performance of a simulation.  In general these units are not essential
to producing scientific results, but provide information to the user about hardware usage and software efficiency.
The Simulation unit
is of particular significance; it defines how a FLASH application will
be built and executed. It also provides initial conditions and the
simulation-specific runtime parameters for the application.  The
Simulation unit has been designed to enable customization of the FLASH
code for specific applications without modifying other units, as
explained in Section~\ref{sec:SimulationUnit}. Additional details on
the unit architecture in general is provided in
Section~\ref{sec:UnitArchitecture}.

\subsection{Configuration Layer}
\label{sec:Configuration}

FLASH implements its inheritance, extensibility, and object-oriented
approach through its configuration layer. This layer consists of a
collection of text {\em Config} files that reside at various
levels of the code organization, and the {\em setup} tool which
interprets the Config files. The two primary functions of
this layer are to configure a single application from the FLASH
source tree, and to implement inheritance and customizability in the
code. The Config files for a unit
contain directives that apply to everything at, or below, that
hierarchical level, and
describe its dependencies as well as
variables and runtime parameters requirements. The setup tool
parses the relevant Config files, starting with the one for the
Simulation unit described in Section~\ref{sec:SimulationUnit}.
Dependencies are recursively resolved to configure individual units needed for the
application. Remember that each application requires different sections of
code and produces a distinct executable. This method of configuration
avoids an unnecessarily large binary and memory footprint, as only the
needed sections of code are included. It also enables extensibility,
since the inclusion of a new unit, or a new implementation of a unit, need
become known only to the Config file of the specific problem setup in the
Simulation unit.

\begin{figure}[htbp]
\begin{center}
 {\subfigure[Config for Simulation]{\includegraphics[width=2.6in]{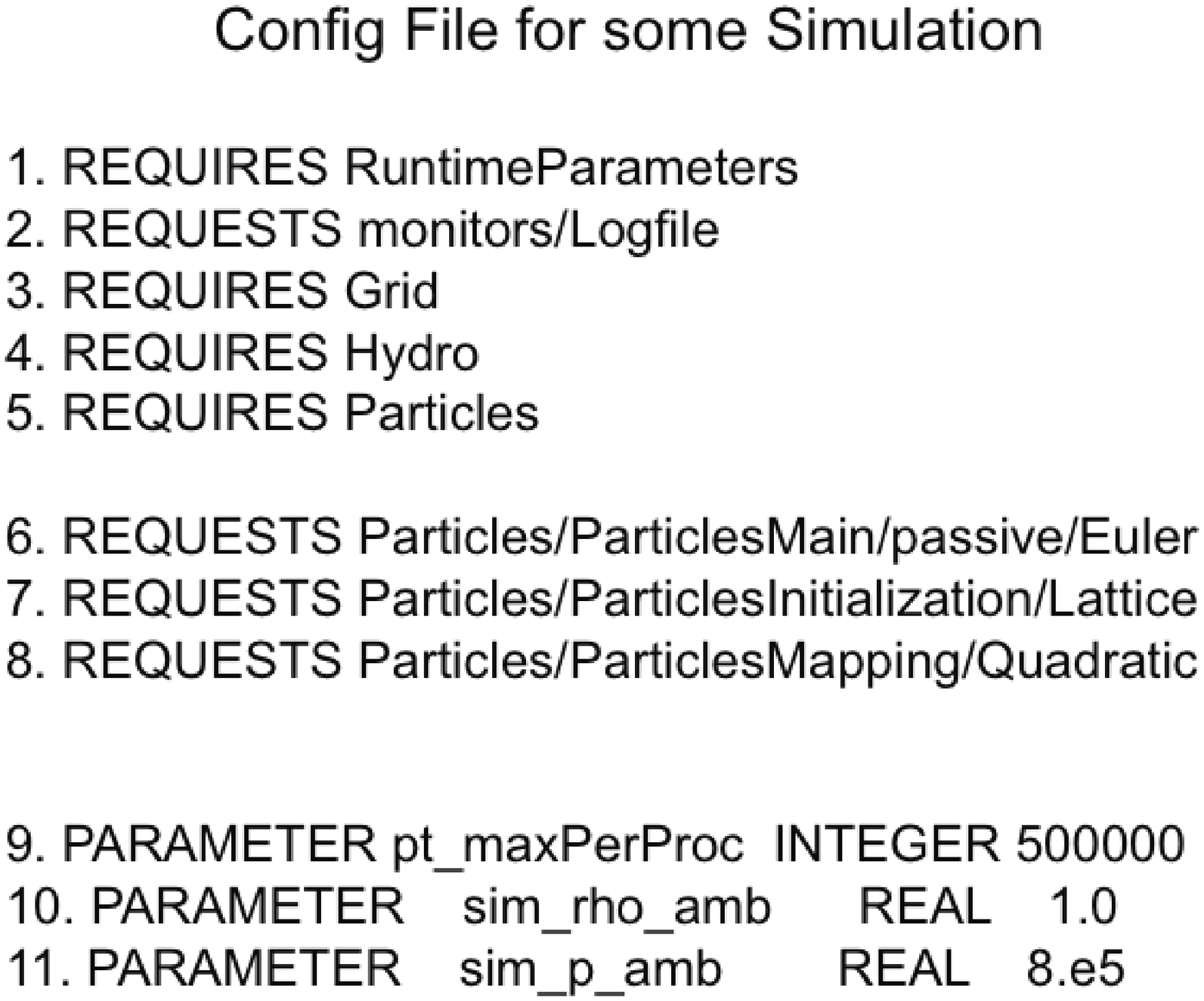}
    \label{fig:config_sim}}}
 {\subfigure[Config for ParticlesMain]{\includegraphics[width=2.6in]{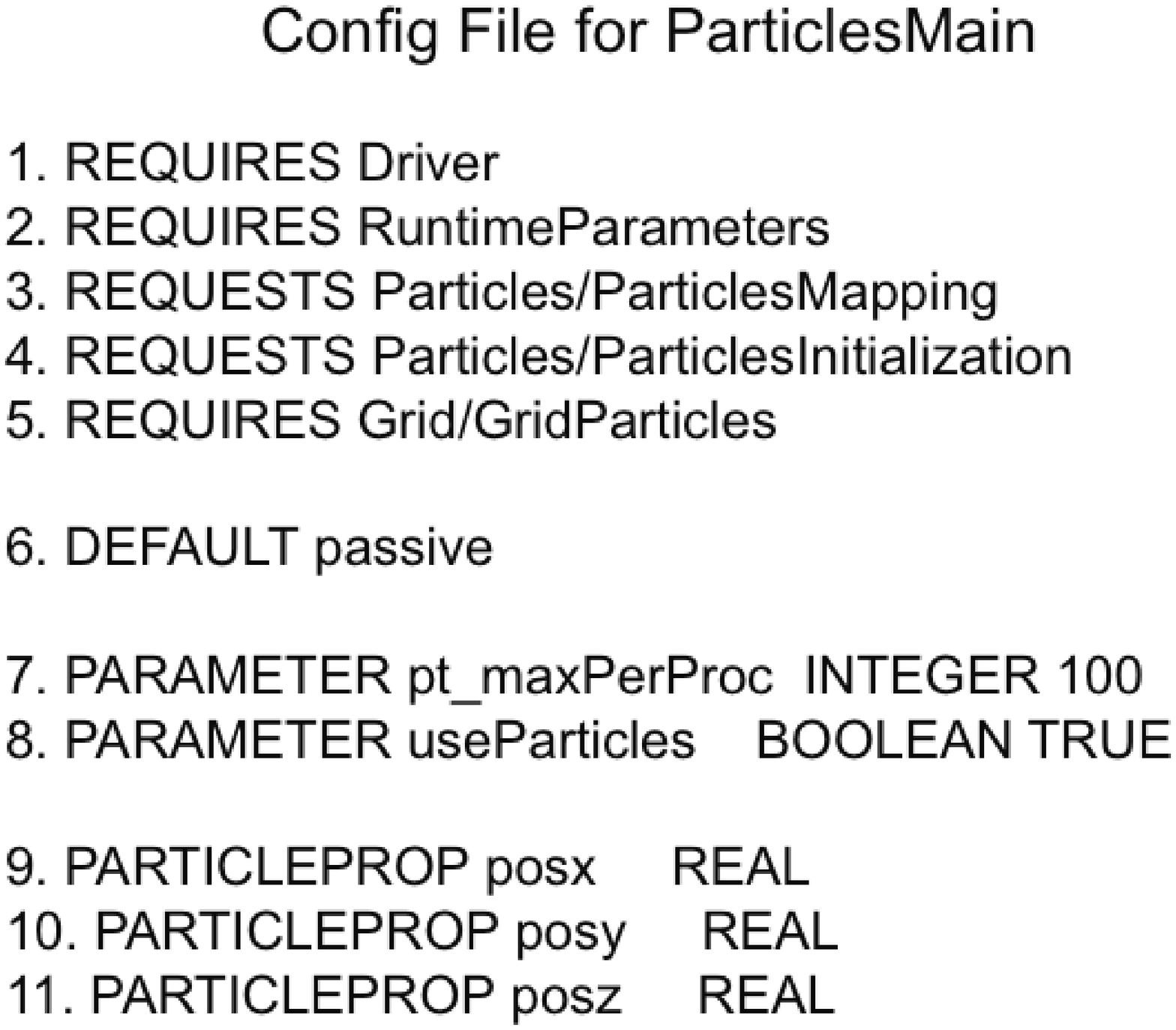}
    \label{fig:config_part}}}
\caption{\label{fig:config} Sections of Sample Config files.}
\end{center}
\end{figure}

Figure \ref{fig:config} shows sections of two sample
Config files, one from the Simulation unit (left panel), and another one from a physics
unit (right panel). In Figure \ref{fig:config_sim}, lines 1 and 3-5
specify units that must be included.
Line 2 specifies a monitoring unit that is requested but may be excluded.
No substitutions are permitted for these units, or their implementations.
In the same file, lines 6-8 specify desirable implementations of subunits.
These subunit implementations will be included if there are no overriding
directives given on the setup command line, but such a directive can cause them
to be either excluded or replaced by another implementation. The
remaining lines in the file pertain to the runtime parameters and
variables. Similarly, in the Config file shown in the right panel of Figure~\ref{fig:config_part},
the first 5
lines specify the required and desirable  units and subunits. Line 6
indicates which implementation of the current unit is to be
included by default,
in this case, an implementation that is found in the ParticlesMain/passive subdirectory.
Again, a directive to the setup tool can replace
this implementation. Note that both the Config files define the
parameter ``pt\_maxPerProc'', along with its default value. Because of
FLASH's inheritance rules, the parameter value in the Simulation Config will be
used in the simulation, which in turn can be overwritten at runtime.

\begin{figure}[htbp]
\begin{center}
 \vspace{-0.5in}
 {\subfigure[Architecture view.]{\includegraphics[width=3.7in]{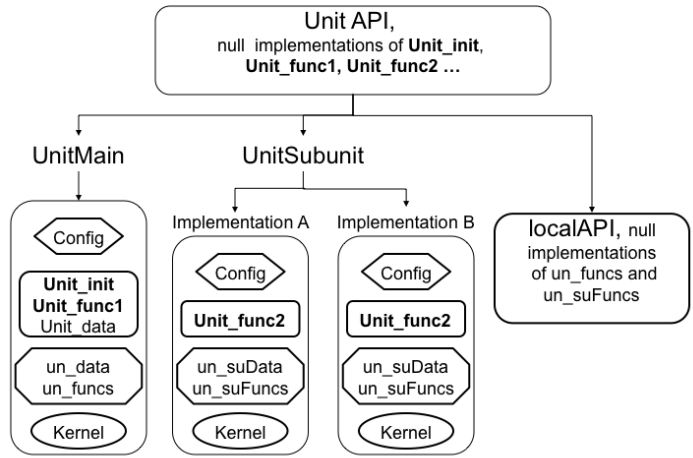}
    \label{fig:units_logical}}}
\\  
 {\subfigure[Unix tree structure view.]{\includegraphics[width=3.2in]{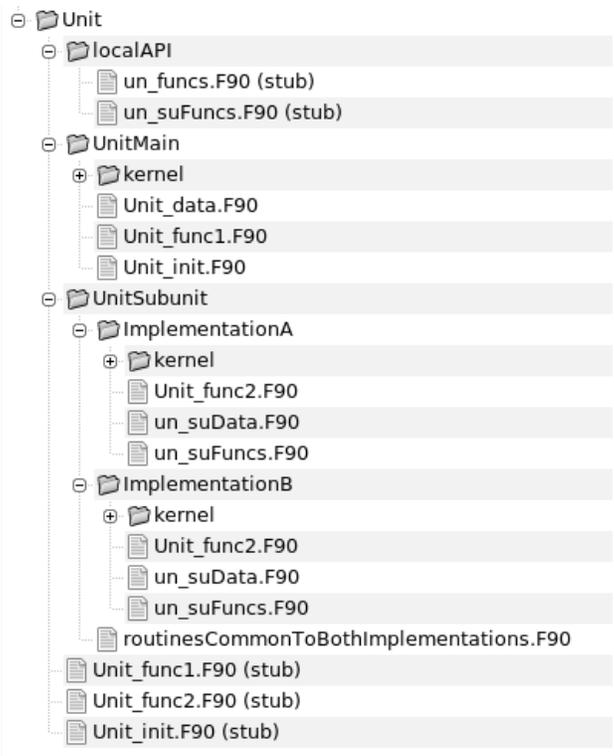}
    \label{fig:units_Unix}}}
\caption{\label{fig:units} Architecture of Units, Subunits, and local API for FLASH.}
\end{center}
\end{figure}

FLASH's approach of using the Unix directory structure with text annotations in the Config files to implement inheritance and other object oriented features has the triple advantage of
being simple, extensible, and completely portable.
Figure \ref{fig:units} shows an example unit and its corresponding Unix directory organization.
The unit has two subunits: one with a single implementation, and another one with two alternative implementations.
The top section of Figure~\ref{fig:units_logical} shows the logical architecture of the unit,
while the bottom section of Figure~\ref{fig:units_Unix}
shows its organization using the Unix directory structure.

\subsection{Data Management}
\label{sec:DataManagement}

In a large multiphysics code with many solvers, management and
movement of data is one of the biggest challenges. Legacy solver codes
rarely address resolving the ownership of data by
different sections of code, a necessity for encapsulation and
modularity. During the first round of modernization in
the second version of FLASH, the data management was centralized into a
separate unit to unravel the legacy code. This technique is also the
data management model followed by SAMRAI \citep{samrai2008}.
The centralized data management extracted all the data from the
individual units, and ensured data coherency by eliminating any
possibility of replication. The main drawback of this approach
was that it gave equal
access to all units for data fetching and modification. Thus a unit
could get mutator access to data that it should never have
modified. The onus was on the developer to find out the scope of each
data item being fetched and to make sure that the scope was not violated. This
responsibility limited the ability to add more functionality to the
code to those who knew the code very well,   a serious handicap to
extensibility.

FLASH Version 3 takes the next and final step in modularizing data
management by decentralizing the data ownership. Every data item in
the code belongs to exactly one unit. The owner unit has complete
control over the scope and modifiability of the data item while the
non-owner units can access or mutate the data only through the owner unit's
API functions. Additionally, the scope of data within a unit can vary.
Thus for example a data item specific to a subunit is visible only
to that subunit, while unit scope data is visible to all functions in
the unit.

\subsection{Interactions Between Units}
\label{sec:Interactions}

The interactions between units are governed by both the {\em Driver}
unit and the published APIs of the individual units. The Driver unit is
responsible for initializing all the included units and the meta-data
for the application as a whole. The Driver unit implements the
time-stepping scheme of the application, and hence dictates the order
in which the units are initialized and invoked, and how they interact
with each other. Recall that units have default null implementations, a
feature that allows a comprehensive implementation of the Driver unit.
Once a unit is invoked by the driver, it can also
interact with other units through their API. The Driver unit also
cleanly closes the units and the application when the run is complete.

\section{Unit Architecture}
\label{sec:UnitArchitecture}
Of the four cornerstones of the FLASH architecture, the unit structure is the most
complex. Unit architecture separates the computational kernel from the
public interfaces, and controls the scope of various data items owned by
the unit. A detailed description of the unit architecture is therefore critical
to understanding the overall structure and software methodology of
the FLASH code.  Subunits are an important and novel feature of the unit architecture detailed below. In addition to the unit architecture, we also describe
some of the infrastructure units and the Simulation unit, since
these play an important role in the code architecture.

The unit itself has three layers. The outer layer, the API, defines the
full functionality of the unit. A unit's API can be viewed as having two sections: one for making its
private data available to the other units, and another which defines
its capabilities for modifying the state of the simulation.
The inner layer of the unit is known as the kernel, and implements
the full functionality. The middle layer implements the
architecture, and acts as conduit between the outer and inner
layers. It hides the knowledge of the FLASH framework and unit
architecture from the kernel, and vice-versa, by providing wrappers
for the kernel. The wrapper layer thus facilitates the import of
third party solvers and software into FLASH.
To include a new third party algorithm, additional wrappers would
be implemented in the middle layer to interface between the
already published API and the new functionality.

\subsection{Subunits}
\label{sec:Subunits}
Units can have one or more subunits which are groupings of
self-contained functionality.  The concept of subunits is new in FLASH version 3.
It was developed to constrain the complexity of the code architecture, and to
minimize the fragmentation of code units, which would result in proliferation of data access functions. In particular the concept of subunits formalizes the selective use of a subset of a unit's functionality, and the possibility of multiple alternative implementations of the same subset. The wrapper layer in the unit architecture starts with the definition of subunits. Subunits
implement disjoint subsets of a unit's API, where none of the subsets
can be a null set. The union of all subsets constituting various
subunits must be exactly equal to the unit API.
Every unit has at least a {\em Main} subunit that implements the
bulk of the unit's functionality, including its initialization.
The Main subunit is also the custodian of all the unit-scope data.
The wrapper layer arbitrates on locating functions common to many alternative
implementations of subunits, such that code duplication is minimized
and flexibility is maximized.

\begin{figure}[htbp]
\begin{center}
\includegraphics[width=5.4in]{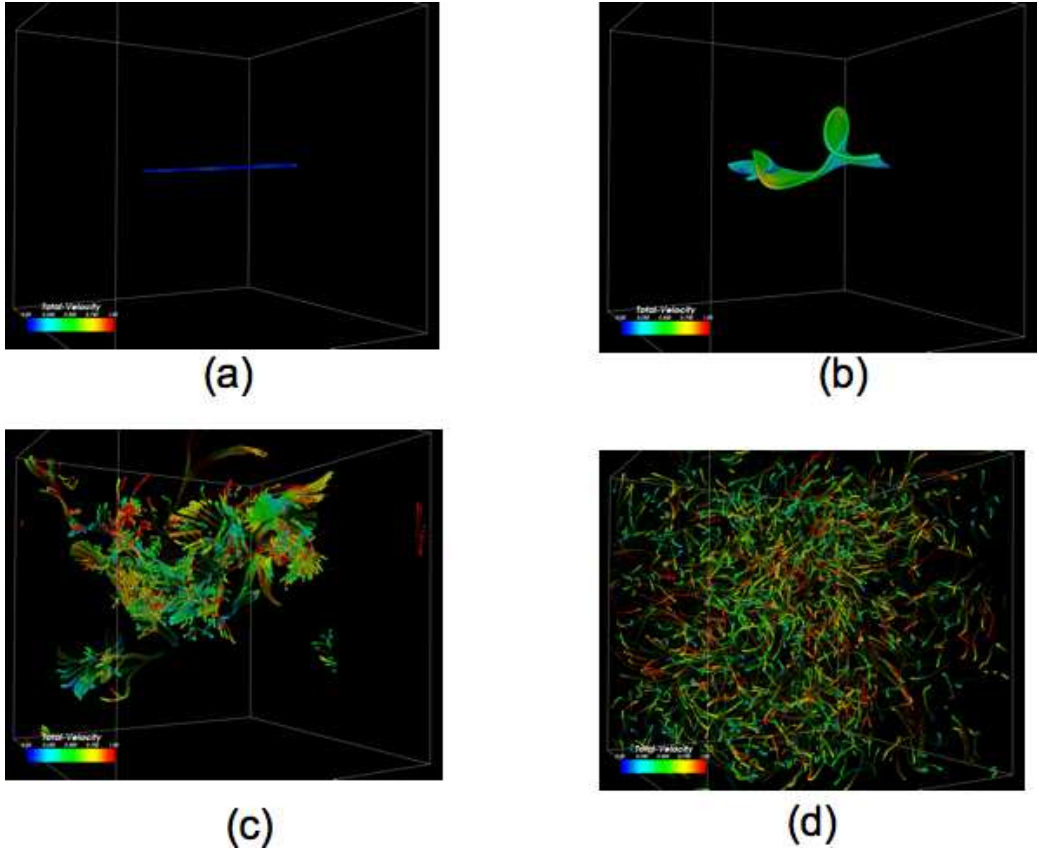}
\caption{\label{fig:particles} Images of Lagrangian tracer particles' movement with advance
in time evolution. The snapshots are taken at times (a) T=0, (b) T=0.75,
(c) T=1.75 and finally (d) T=4.25 seconds. The simulation was done on
32,768 nodes of the IBM BG/L machine at Lawrence Livermore National Laboratory, with $1856^3$ grid points and more than 16
million particles.}
\end{center}
\end{figure}

The use of the subunit concept is best illustrated with an example of interdependencies between the \textit{Grid} unit, which manages the Eulerian mesh, and the \textit{Particles} unit.
The discretized mesh in FLASH is composed of a collection of blocks, where
individual blocks span a section of the domain, and all the blocks
taken together cover the entire domain. In parallel environments,
domain decomposition maps one or more blocks to each processor
participating in the simulation.
Particles may be massless and passive, used to track the Lagrangian features of the
simulation, or active particles with mass which can affect gravitational fields. While individual elements (zones and grid points) of the Eulerian mesh stay at the same physical location in the domain throughout the evolution, the Lagrangian elements (particles) move with the motion of the fluid. The motion of Lagrangian particles relative to the underlying Eulerian mesh is best illustrated with snapshots of a set of particles at different times during evolution. Figure \ref{fig:particles} shows the positions of a small subset of particles at different stages of evolution in a weakly compressible turbulence simulation using a uniform grid \cite{Fisher2008}. Here, because the mesh does not change with time, the Eulerian elements are stationary in the physical domain at all times, while  the narrow line of Lagrangian elements has spread all over the domain in the same timeframe.

FLASH has four distinct subsets of functionality related to particles, each of which can have
multiple alternative implementations.  The current FLASH release provides three
implementation methods for initial distribution of particles,
four methods of mesh/particle mapping, two types of gravitational field interaction,
and seven methods of time integration.
This level of complexity is not limited to the Particles unit.
The time integration of particles can result in their migration between
physical regions served by different processors.
Similarly, regridding of the active mesh may require migration of particles.
These particle-related movements
are best handled by the Grid unit since it knows the topology of the Eulerian mesh,
thereby retaining encapsulation of alternative unit implementations.

If FLASH were to solely follow the unit
model of architecture described in Section~\ref{sec:Unit}, then separate units for particles
distribution, mapping, integration and migration would be needed.
Each of these units would need access to large amounts of data in the other units, thereby requiring many accessor-mutator functions.  Therefore, the addition of subunits is a major feature of the \FLASHthree architectural improvements.
The concept of subunits very elegantly solves both the problems of data access and
unit fragmentation through the introduction of a level of hierarchy in
the unit's architecture. Thus in the Particles unit the
ParticlesInitialization and ParticlesMapping subunits
respectively deal with the initial spatial distribution and with mappings
to and from the Eulerian grid, while the ParticlesMain unit keeps
the unit scope data and implements time integration methods. Each subunit can have several
alternative  implementations. Hence, subunits not only organize a unit into
distinct functional subsets that can be selectively turned off, but
also expand the flexibility of the code since implementations of
different subunits can permute with each other and therefore can be
combined in many different ways.

\subsection{Lateral Data Movement}
\label{sec:Lateral}
In addition to the subunits level functionality, the other major
challenge posed by the interaction between solvers for multiphysics
simulations is the need for lateral data movement, which makes
resolution of data ownership and encapsulation extremely
difficult.  For instance, the calculation of the hydrodynamics
equations is dependent upon the equation of state, and if gravity is
included in the simulation, upon gravitational acceleration. Similarly,
within the hydrodynamics calculation, there is a need to reconcile the
fluxes at a global level when adaptive meshing is being used. All of
these  operations require access to data which is owned by different
units.  Though version 2.5 of FLASH with its centralized database did not have some of these difficulties, it did not resolve data ownership, and did not achieve encapsulation. \FLASHthree's
solution to this challenge is to provide interfaces that allow for
transfer back and forth between units, so that data can be accessed through
argument passing by reference. The challenge is then reduced to
arbitration between units as to which one is best suited to implement
the needed functionality.

\begin{figure}[htbp]
\begin{center}
 {\subfigure[Particle advancement in time.]{\includegraphics[height=1.8in]{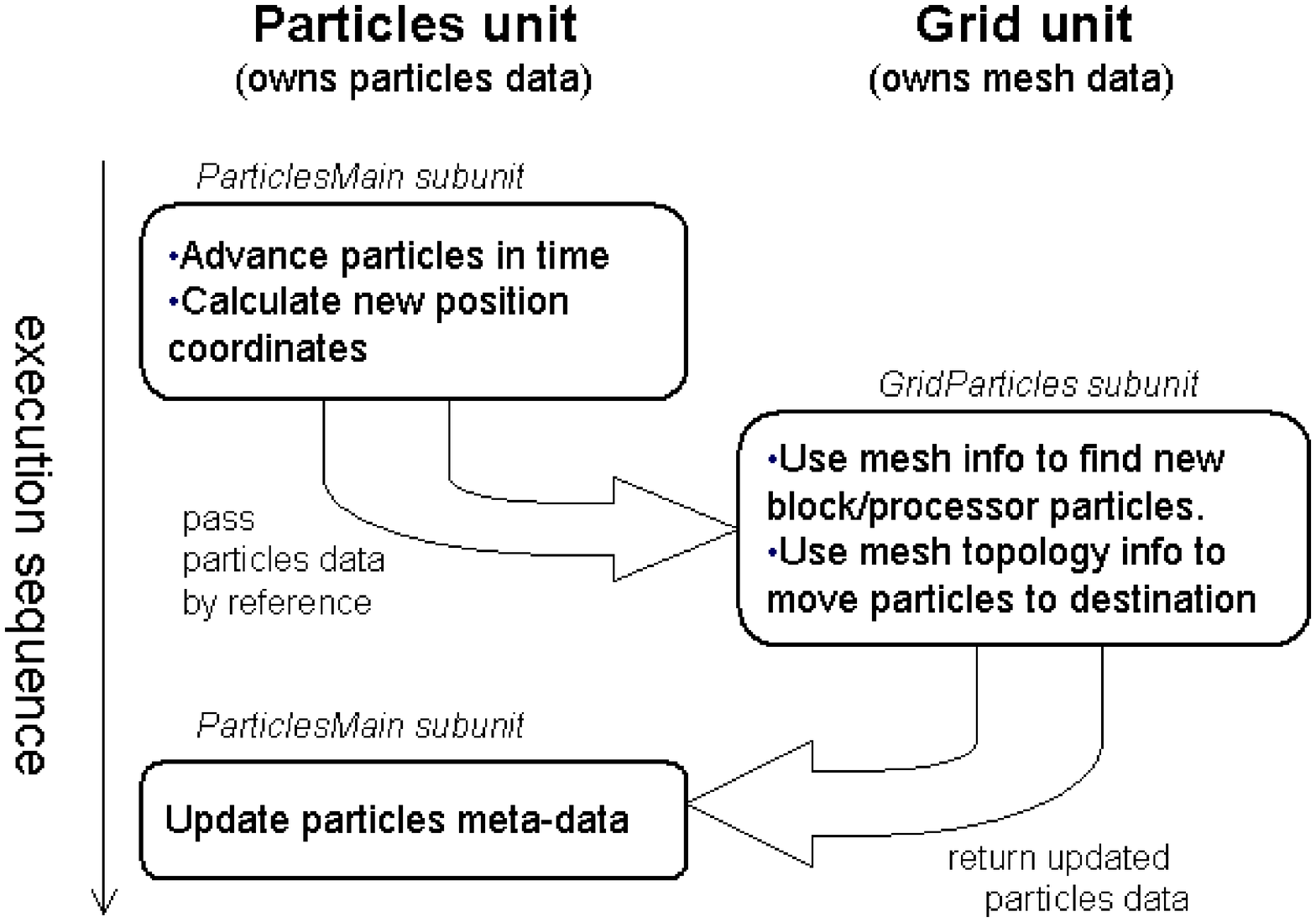}
    \label{fig:lateral_particles}}}
 {\subfigure[Mesh refinement.]{\includegraphics[height=1.8in]{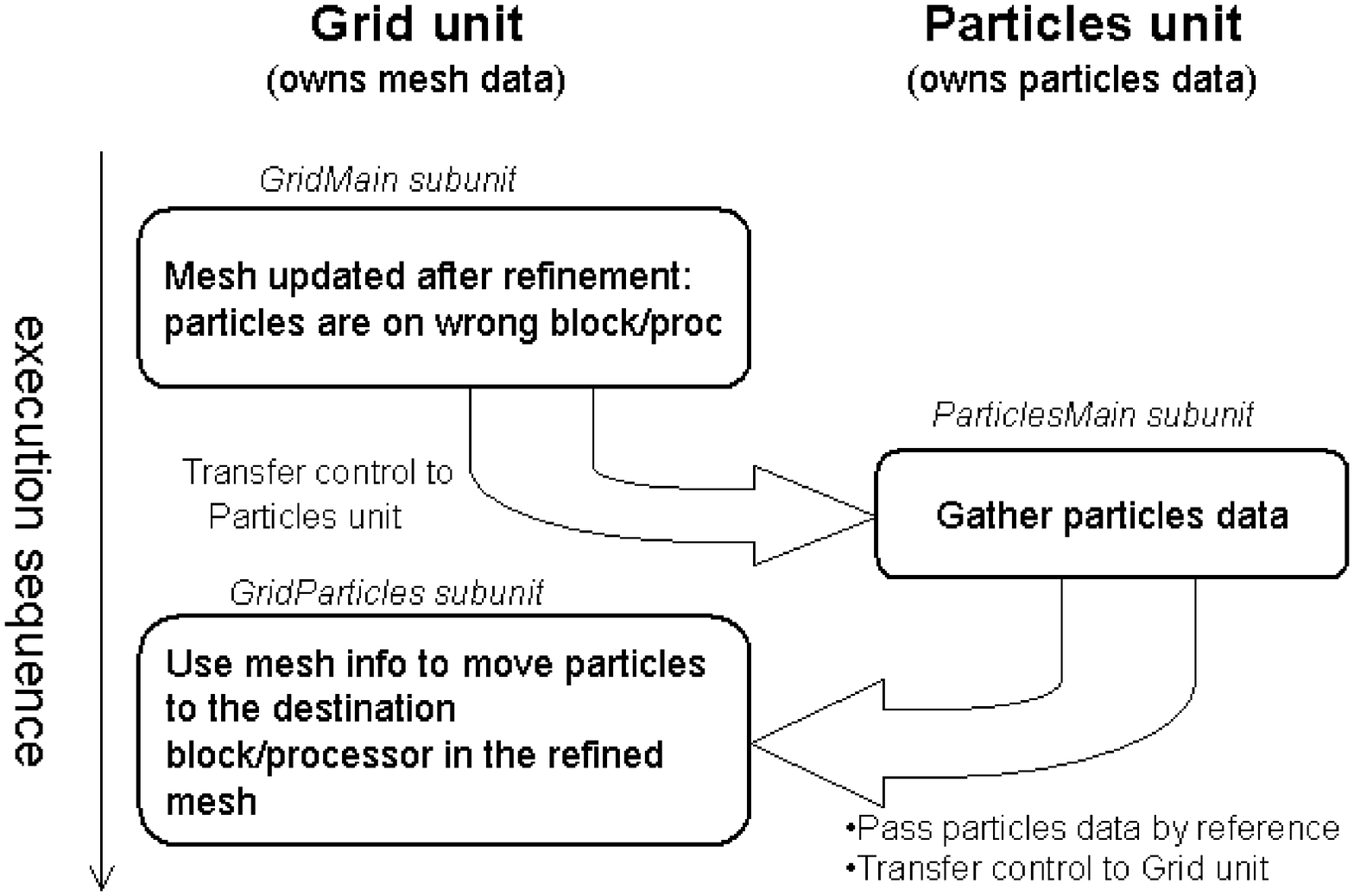}
    \label{fig:lateral_grid}}}
\caption{\label{fig:lateral} Lateral Data Movement during two different algorithmic steps.}
\end{center}
\end{figure}

Figure~\ref{fig:lateral} shows examples of
lateral data movement between the Particles unit and the Grid unit.
The left panel of Figure~\ref{fig:lateral_particles} shows the flow of
execution, starting in the Particles unit, as particles change their
physical position due to time integration.  Some of the new positions
in the Eulerian mesh
may be on different processors.  The movement of particles to the
appropriate processor is best carried out by handing control, along
with the particles data, to the Grid unit because of its knowledge of
the mesh layout.
Once it has moved the particles appropriately, the Grid unit returns the data and control back to
the Particles unit.  The
right panel of Figure~\ref{fig:lateral_grid} shows movement between the
same two units where the example operation starts in the Grid unit.
When using AMR, the mesh regridding
operation changes the mapping of blocks to processors. In reorienting
themselves to the new mesh, the particles have to move among
processors. Because the particles' data structures are not accessible
to the Grid unit,  the control is temporarily transferred to the Particles unit, which passes the particles' data by reference to the Grid unit for redistribution. Both examples preserve data encapsulation and ownership without compromising the performance.

\subsection{Infrastructure Units}
\label{sec:InfrastructureUnit}

The infrastructure units in FLASH are responsible for discretization
of the physical domain; reading, writing, and
maintaining the data structures related to the simulation data; and
other housekeeping tasks such as handling physical constants and
runtime parameters. Of these, the most extensive responsibilities lie
with the Grid unit, which manages the discretized mesh, and the input/output
IO unit, which reads and writes the data. These two units are
also unique in that they share their data with each other; this
exception to unit encapsulation is allowed for performance reasons.
Here we describes these two
units briefly (further discussion is found in \citep{flash2008} and \citep{dubey2008b}).

The Grid unit is the custodian of all the data structures related to
the physical variables necessary for advancing the simulations.
Every discrete point in the mesh is associated with a
number of physical variables, logical and physical coordinates, and an
indexing number.  On each processor, meta-data exists,
such as the location in the physical domain and the
number of discretization points per parallel grouping.
\FLASHthree has two different Grid
implementations:  a simple grid uniform in space and a block-structured adaptive oct-tree mesh.
If Adaptive Mesh Refinement is being used, blocks are created,
destroyed, and distributed dynamically, and different blocks exist at varying levels
of resolution, all of which must be tracked by the unit. The Grid unit
is also responsible for keeping the physical variables consistent
throughout the simulation. For example, when two adjacent blocks are
at different resolutions, interpolation and prolongation
ensure that conservation laws are not violated.
Hence, the Grid unit is the most complex and extensive unit in the code,
and most of the scaling performance of the code is determined by the
efficiency of its parallel algorithms.

In FLASH, more than 90\% of the reading or writing of data to the disk
is controlled by the IO unit.
FLASH outputs data for checkpointing
and analysis. The checkpoints save the complete state of the
simulation in full precision so that simulations can transparently
restart from a checkpoint file.
The analysis data is written in many formats. The
largest of these are the plotfiles, which record the state of the physical variables.
Quantities integrated over the entire
domain are written from the master processor into a simple text file.
The only input controlled by the IO unit is the reading of
checkpoint files. Other forms of input, such as reading in a table of
initial conditions needed by a specific simulation, are managed by the
unit in question.
FLASH is one of the relatively few applications
codes that have support for multiple IO libraries, such as HDF5 \citep{HDF5url} and
parallel netCDF \citep{Li2003,HDF5comparison}, where all processors can write data to a single shared file.

\subsection {Simulation Unit}
\label{sec:SimulationUnit}
The {\em Simulation} unit
effectively
defines the scientific application.  Each subdirectory in the
Simulation unit contains a different application, which can be viewed
as a different implementation of the Simulation unit. This unit also provides a mechanism
by which users can customize any part of their application without having to
modify the source code in any other unit.

An application can assume very specific knowledge of units it
wants to include and can selectively replace functions from other
units with its own customized ones by simply placing a different
implementation of the function in its Simulation subdirectory.
At configuration time, the arbitration
rules of the setup tool cause an implementation placed
in the simulation unit to override any other implementation of that
function elsewhere in the code.
Similarly, the
simulation unit can also be aware of the runtime parameters defined in
other units and can reset their default values. Additionally, FLASH
does not limit applications to the functionality distributed with the
code; an application can add functionality by placing its implementation in
the Simulation subdirectory. The setup tool has the capability to include any new
functionality thus added at configuration time, without any prior
knowledge of the functionality. Accordingly, by allowing great flexibility to the Simulation
unit, FLASH makes it possible for users to quickly and painlessly
customize the code for their applications. A typical use of this
flexibility is in user-defined boundary conditions that may not have
standard support in FLASH. Another frequently customized functionality
is control of refinement when using the AMR adaptive grid mode.

\section{Code Maintenance}
\label{sec:CodeMaintenance}
While a clear architecture design is the first step in producing a useful code,
the FLASH code is not static and continues to develop based on internal
pressures and external requests and collaborations.
As the code gains maturity, regular testing and maintenance become crucial.
Maintenance of the FLASH code is assisted by
guidelines for all stages in the code lifecycle,  some of which are
enforced and others are strongly encouraged.

\subsection{Unit Test Framework}
\label{sec:UnitTest}
In keeping with good software practice, \FLASHthree incorporates a
unit test framework that allows for rigorous testing and easy
isolation of errors.
The implementation of a new code unit or subunit
is usually accompanied by the creation of one or more corresponding
unit tests. Where possible, the unit tests compare numerical results
against known analytical or semi-analytical solutions which isolate
the new code module.

The components of the unit test reside in two
different places in the FLASH source tree. One is a dedicated path in
the Simulation unit, where the specific unit test acts as an ordinary Simulation.
The other is a subdirectory called unitTest, located within
the hierarchy of the corresponding unit, which implements
the actual test and any helper functions it may need.  These functions
have extensive access to the internal data of the unit being tested.
By splitting the unit test into two locations in the source tree,
unit encapsulation is maintained.

\begin{figure}[htbp]
\begin{center}
\includegraphics[width=5in]{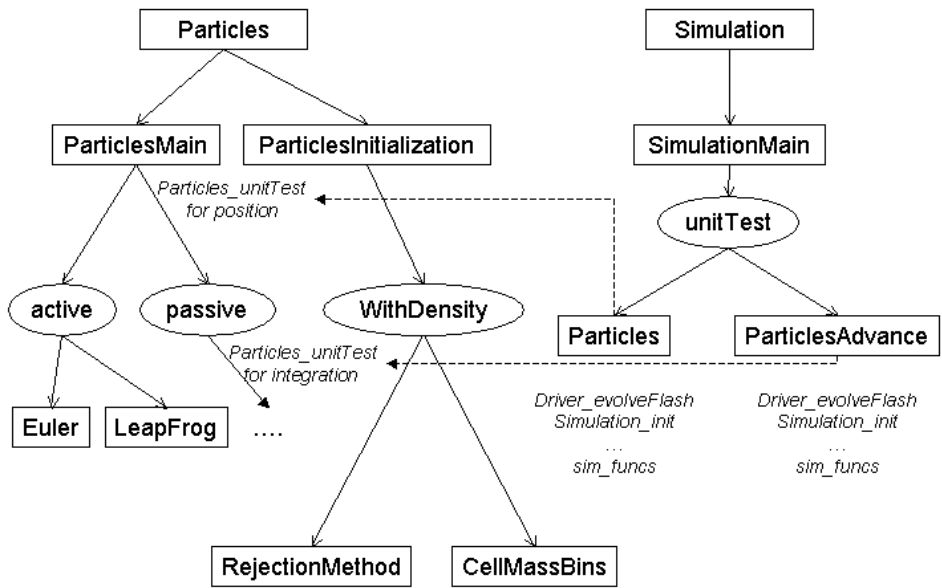}
\caption{\label{fig:unitTest} The unit test framework underlying the FLASH source tree.
Unit tests are split into drivers located in a subdirectory of the Simulation unit and implementation routines within the
relevant unit being tested.  {\em Files} are shown in italics.  Dotted lines indicate the coupling between the two units.}
\end{center}
\end{figure}

Figure~\ref{fig:unitTest} illustrates the split implementation of the unit test with an example.
The figure shows relevant sections of the Particles  and Simulation units in the FLASH code.  The example does not represent the full implementation of either unit; it includes only those few sections that best highlight the features of
the unit test framework. In the Simulation unit, there is an organizational directory which houses all the unit tests. Within this directory, there are two unit tests for the Particles unit.
One of the tests verifies the correct movement of the particles after their positions have changed because of either time integration or regridding. The routine implementing this test resides at the top level of the ParticlesMain  subunit.
The other unit test verifies the time integration methods that advance passive particles in time. For this test, the corresponding routine resides in the subdirectory "passive" of  ParticlesMain subunit, where time integration of passive particles is computed.
Figure~\ref{fig:unitTest} also shows the ParticlesInitialization subunit to facilitate clearer understanding of the unit structure and the overlying unit test framework.
The  dotted arrows from the  Simulation unit test to the Particles unit show the coupling between the two units.
The figure also highlights  the flexibility of having alternative implementations of
the same function co-exist at several levels in the source tree.

\subsection{Documentation}
\label{sec:Documentation}
FLASH's clean architecture is well documented, which enables
easy extention by external contributors \citep{flash2008}.
For all routines defining the interface of a unit, a well documented
header is a code requirement. The developers are also strongly
encouraged to include extensive in-line documentation in addition to a
header describing each routine  they implement. FLASH uses Robodoc
\citep{robodoc2007,robodoc2005}
for automatic generation of documentation from internal
headers. Compliance with code regulations such as documentation and
good coding practice is checked through scripts that run nightly.

In addition,  rapidly executing example problems are provided in
the public release of FLASH.
Availability of a collection of example problems
that a first-time user can set up and run in an hour or less has been
cited as one of the more attractive features of FLASH in a code survey (see Section~\ref{sec:UserSurvey}).
FLASH comes with a  User's
Guide, on-line howtos, on-line quick reference tips,
and hyperlinks to full descriptions with examples of all the API routines that
form the public interfaces of various units \citep{flash2008a,flash2008}.
All of these user-assistance
components are available on-line, as is the current release.  In addition,
there is an active email User's Group where support questions are addressed
by both developers and knowledgeable active users.

\section{User Survey}
\label{sec:UserSurvey}

The FLASH Code has attracted a wide range of users and has become a
premier community code preeminent in, but not limited to, the astrophysics
community. Many users cite FLASH's capabilities, ease of use,
scalability, modularity, and extensive documentation as the key
reasons for their use of FLASH.  A code survey performed in 2005,
followed by another in 2007, found that the close to three hundred
responding users utilize the code in three
major ways.  The  first group (approximately 41\%)
uses FLASH as a primary research tool
for a broad range of application areas, including high-energy
astrophysics, cosmology, stars and stellar evolution, computational
fluid dynamics (CFD), and algorithm development.
The second group of users ($\approx 9\%$) employ the FLASH
code for verification and validation (V\&V).  These users primarily attempt to
compare FLASH to other codes or use FLASH as a benchmark. Still others
in this V\&V group port FLASH to new machines to test compilers,
libraries, and performance.  Finally, the third group ($\approx 25\%$) uses FLASH as a
sample code or for educational purposes.

\begin{figure}[http]
\begin{center}
\includegraphics[scale=0.55]{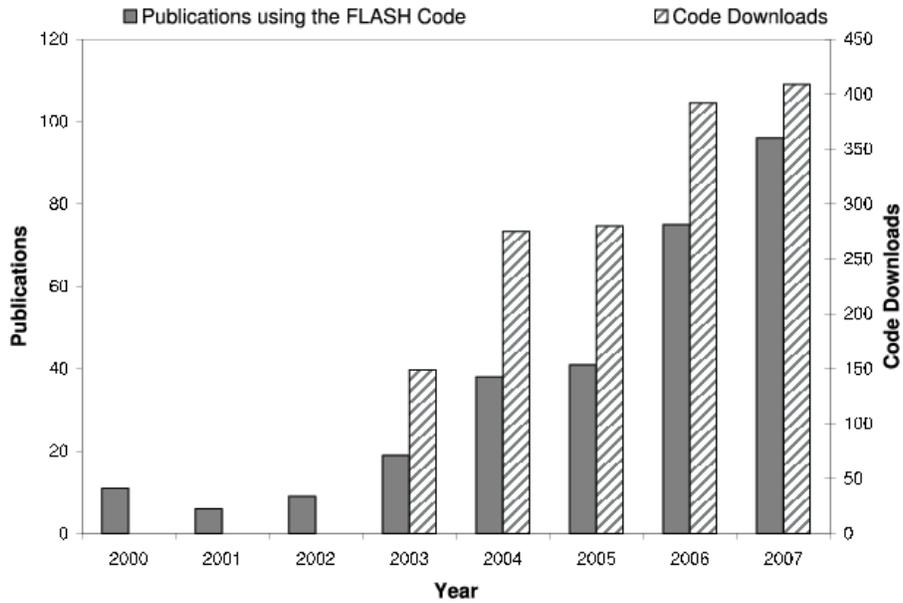}
\caption{Yearly number of publications in which the FLASH code was used
(left dark bars) and FLASH downloads (right striped bars).
The jump in downloads in 2006 followed the release of the alpha version of \FLASHthree, the
new version of the code. }
\label{Fig:FlashDownloadsPapers}
\end{center}
\end{figure}

\begin{figure}[htbp]
\begin{center}
\includegraphics[scale=0.55 ]{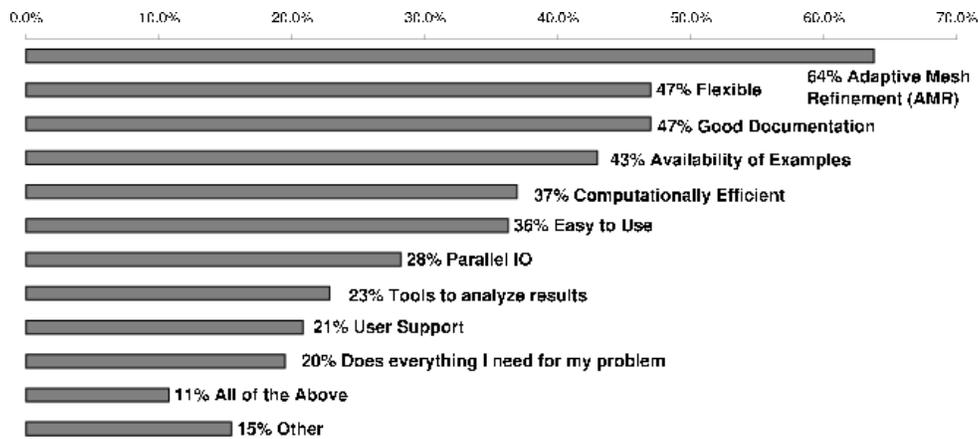}
\caption{Results from a FLASH users survey in 2007:  Reasons cited for FLASH usage.}
\label{Fig:WhyUsingFlash}
\end{center}
\end{figure}

The results of the survey clearly indicate that FLASH enjoys wide
acceptance among researchers from many fields. By 2007, FLASH had been
downloaded more than 1700 times and used in more than 320 publications,
by both Center members and external users.  Figure
\ref{Fig:FlashDownloadsPapers} shows that both the number of code
downloads and the number of publications has steadily grown as the
code has matured. Figure \ref{Fig:WhyUsingFlash} shows that while the
presence of adaptive mesh refinement is the top reason cited for using
FLASH, it is the only one in the top six reasons that relates to the
capabilities of the code. The remaining five top reasons pertain to the code
architecture and its software process. These reasons include flexibility,
ease of use, and performance, thus vindicating the architectural
choices of FLASH.

\section{Acknowledgments}
We wish to thank all the past contributors to the FLASH code.
The software described in this work was in part developed by the DOE-supported ASC / Alliance Center for Astrophysical Thermonuclear Flashes at the University of Chicago under grant B523820.

\thanks{The software described in this work was in part developed by the DOE-supported ASC / Alliance ASC/Flash Center at the University of Chicago under grant B523820.}

\newpage  

\noindent
{\bf Figure 1 Caption:} \\
Sections of Sample Config files.
\\
\noindent
{\bf Figure 2 Caption:} \\
Architecture of Units, Subunits, and local API.
\\
\noindent
{\bf Figure 3 Caption:} \\
Images of Lagrangian tracer particles' movement with advance
in time evolution. The snapshots are taken at times (a) T=0, (b) T=0.75,
(c) T=1.75 and finally (d) T=4.25 seconds. The simulation was done on
32,768 nodes of the IBM BG/L machine at Lawrence Livermore National Laboratory, with $1856^3$ grid points and more than 16
million particles.
\\
\noindent
{\bf Figure 4 Caption:} \\
Lateral Data Movement during two different algorithmic steps.
\\
\noindent
{\bf Figure 5 Caption:} \\
The unit test framework underlying the FLASH source tree.
Unit tests are split into drivers located in a subdirectory of the Simulation unit and implementation routines within the
relevant unit being tested.  {\em Files} are shown in italics.  Dotted lines indicate the coupling between the two units..
\\
\noindent
{\bf Figure 6 Caption:} \\
Yearly number of publications in which the FLASH code was used
(left dark bars) and FLASH downloads (right striped bars).
The jump in downloads in 2006 followed the release of the alpha version of \FLASHthree, the
new version of the code.
\\
\noindent
{\bf Figure 7 Caption:} \\
Results from a FLASH users survey in 2007:  Reasons cited for FLASH usage.

\begin{thebibliography}{10}
\expandafter\ifx\csname url\endcsname\relax
  \def\url#1{{\tt #1}}\fi
\expandafter\ifx\csname urlprefix\endcsname\relax\def\urlprefix{URL }\fi
\providecommand{\eprint}[2][]{\url{#2}}

\bibitem[Antypas et al.(2006)]{antypas2006}
Antypas, K.B., Calder, A.C., Dubey, A., Gallagher, J.B.,
Joshi, J., Lamb, D.Q., Linde, T., Lusk, E.L., Messer, O.E.B.,
Mignone, A., Pan, H., Papka, M., Peng, F., Plewa, T.,
Riley, K.M., Ricker, P.M., Sheeler, D., Siegel, A.,
Taylor, N., Truran, J.W., Vladimirova, N., Weirs, G., Yu D., and Zhang, J. (2006).
\newblock FLASH: Applications and Future.
\newblock \emph{Parallel Computational Fluid Dynamics 2005: Theory and Applications} 235+.


\bibitem{armstrong2006}
Armstrong, R., Kumfert, G., McInnes, L., Parker, S., Allan, B., Sottile, M.,
  Epperly, T., and Dahlgren, T. (2006).
\newblock The {CCA} component model for high-performance scientific computing.
\newblock \emph{Concurrency and Computation: Practice and Experience} 18(2), 215--229.

\bibitem{flash2008a}
{ASC Flash Center} (2009).
\newblock {FLASH} code support.
\newblock \url{http://flash.uchicago.edu/website/codesupport/}.

\bibitem{flash2008}
{ASC Flash Center} (2009).
\newblock {FLASH} user's guide.
\newblock
\url{http://flash.uchicago.edu/website/codesupport/flash3_ug_3p2.pdf}.

\bibitem{Calder2002a}
{Calder}, A.~C., {Fryxell}, B., {Plewa}, T., {Rosner}, R., {Dursi}, L.~J.,
  {Weirs}, V.~G., {Dupont}, T., {Robey}, H.~F., {Kane}, J.~O., {Remington},
  B.~A., {Drake}, R.~P., {Dimonte}, G., {Zingale}, M., {Timmes}, F.~X.,
  {Olson}, K., {Ricker}, P., {MacNeice}, P., and {Tufo}, H.~M. (2002).
\newblock On validating an astrophysical simulation code.
\newblock \emph{Astrophysical Journal, Supplement} 143, 201--229.

\bibitem{samrai2008}
Center for Applied Scientific Computing (CASC) (2007).
\newblock {SAMRAI} structured adaptive mesh refinement application
  infrastructure.
\newblock \url{https://computation.llnl.gov/casc/SAMRAI/}.
\newblock {CASC, Lawrence Livermore National
  Laboratory}.

\bibitem{HDF5comparison}
Chilan, C., Yang, M., Cheng, A., and Arber, L. (2006).
\newblock Parallel {I/O} performance study with {HDF5}, a scientific data
  package.
\newblock
  \url{http://www.hdfgroup.uiuc.edu/papers/papers/ParallelIO/ParallelPerformance.pdf}.

\bibitem{chombo2009}
  Colella, P., Graves, D.~ T., Keen, N.~D., Ligocki,  T. ~J., Martin, D. ~F.,
  McCorquodale, P. ~W., Modiano, D.,  Schwartz, P.~O., Sternberg, T.~D., and Van Straalen, B. (2009).
  \newblock {Chombo} Software Package for AMR Applications, Design Document,
 \newblock  \url{https://seesar.lbl.gov/ANAG/chombo}.

\bibitem{dubey2008b}
Dubey, A., Reid, L., and Fisher, R. (2008).
\newblock Introduction to {FLASH 3.0}, with application to supersonic
  turbulence.
\newblock \emph{Physica Scripta} 132, 014046.

\bibitem{Fisher2008}
 Fisher, R.,  Abarzhi, S.,  Antypas, K., Asida, S.M., Calder,A.C., Cattaneo, F., Constantin, P., Dubey, A., Foster,I., Gallagher, J.B., Ganapathy, M.K., Glendenin, C.C., Kadanoff, L., Lamb, D.Q., Needham, S., Papka, M., Plewa,T., Reid, L.B., Rich, P., Riley, K., Sheeler, D.(2008).
\newblock Terascale Turbulence Computation on {BG/L} Using the {FLASH3} Code.
\newblock \emph{IBM Journal of Research and Development} 52(1/2), 127--137.

\bibitem{Fryxell2000}
{Fryxell}, B., {Olson}, K., {Ricker}, P., {Timmes}, F.~X., {Zingale}, M.,
  {Lamb}, D.~Q., {MacNeice}, P., {Rosner}, R., {Truran}, J.~W., and {Tufo}, H.
  (2000).
\newblock {FLASH}: An adaptive mesh hydrodynamics code for modeling
  astrophysical thermonuclear flashes.
\newblock \emph{Astrophysical Journal, Supplement} 131, 273--334.

\bibitem{Gardiner2005}
Gardiner, T.~A. and Stone, J.~M. (2005).
\newblock An unsplit {Godunov method} for ideal {MHD} via constrained
  transport.
\newblock \emph{J. Computational Physics} 205(2), 509--539.

\bibitem{hornung2002}
Hornung, R. and Kohn, S. (2002).
\newblock Managing application complexity in the {SAMRAI} object-oriented
  framework.
\newblock \emph{Concurrency and Computation: Practice and Experience} 14(5), 347--368.

\bibitem{Hornung2006}
Hornung, R.~D., {Wissink}, A.~M., and {Kohn}, S.~R. (2006).
\newblock Managing complex data and geometry in parallel structured {AMR}
  applications.
\newblock \emph{Engineering with Computers} 22, 181--195.

\bibitem{hovland2003}
Hovland, P., Keahey, K., McInnes, L.~C., Norris, B., Diachin, L.~F., and
  Raghavan, P. (2003).
\newblock A quality of service approach for high-performance numerical
  components.
\newblock In \emph{Proceedings of Workshop on {QoS} in Component-Based Software
  Engineering, Software Technologies Conference}. Toulouse, France.

\bibitem{Ko2005}
Ko, S., Cho, K.~W., Song, Y.~D., Kim, Y.~G., Na, J., and Kim, C. (2005).
\newblock Development of {Cactus} driver for {CFD} analyses in the grid
  computing environment.
\newblock In \emph{Advances in Grid Computing - EGC 2005} vol. 3470, pp. 771--777.

\bibitem{Li2003}
Li, J., Liao, W., Choudhary, A., Ross, R., Thakur, R., Gropp, W., Latham, R.,
  Siegel, A., Gallagher, B., and Zingale, M. (2003).
\newblock Parallel net{CDF}: A high-performance scientific {I/O} interface.
\newblock \emph{Supercomputing, 2003 ACM/IEEE Conference}  39+.

\bibitem{HDF5url}
NCSA (2008).
\newblock {Heirarchical Data Format 5}.
\newblock \url{http://hdf.ncsa.uiuc.edu/HDF5/}.

\bibitem{Oldham2002}
Oldham, J. (2002).
\newblock Scientific computing using {POOMA}.
\newblock \emph{C++ Users Journal} 20(11), 6--23.

\bibitem{O'Shea2005}
{O'Shea}, B.~W., {Bryan}, G., {Bordner}, J., {Norman}, M.~L., {Abel}, T.,
  {Harkness}, R., and {Kritsuk}, A. (2005).
\newblock Introducing {Enzo}, an {AMR} cosmology application.
\newblock In Plewa, T., Timur, L., and Weirs, V. (eds.) Adaptive Mesh
  Refinement -- Theory and Applications. Springer, vol.~41 of {\em Lecture
  Notes in Computational Science and Engineering\/}.

\bibitem{Reynders1996}
Reynders, J., Hinker, P., Cummings, J., Atlas, S., Banerjee, S., Humphrey, W.,
  Karmesin, S., Keahey, K., Srikant, M., and Tholburn, M. (1996).
\newblock {POOMA: A} framework for scientific simulations on parallel
  architectures.
\newblock \emph{Parallel Programming using C++}.

\bibitem{robodoc2007}
Slothouber, F. (2007).
\newblock Automating software documentation with \newline {ROBODoc}.
\newblock \url{http://www.xs4all.nl/~rfsber/Robo/robodoc.html}.

\bibitem{toth2005}
Toth, G., Sokolov, I., Gombosi, T., Chesney, D., Clauer, C., De~Zeeuw, D.,
  Hansen, K., Kane, K., Manchester, W., Oehmke, R., et~al. (2005).
\newblock {Space Weather Modeling Framework: A} new tool for the space science
  community.
\newblock \emph{J. Geophysical Research} 110, 12--226.

\bibitem{wissink2000}
Wissink, A. and Hornung, R. (2000).
\newblock {SAMRAI: A} framework for developing parallel {AMR} applications.
\newblock In \emph{5th Symposium on Overset Grids and Solution Technology}, Davis, CA,
  pp. 18--20.

\bibitem{robodoc2005}
Worth, D. and Greenough, C. (2005).
\newblock A survey of available tools for developing quality software using
  {Fortran 95}.
\newblock Technical report {RAL-TR-2005}, \emph{{SFTC} Rutherford Appleton
  Laboratory, {SESP} Software Engineering Support Programme}.
\newblock Available at
  \url{http://www.sesp.cse.clrc.ac.uk/html/Publications.html}.

\end{thebibliography}
\end{document}